\documentclass[
]{ceurart}

\usepackage{listings}

\lstdefinestyle{tfmstyle}{
    language=java, 
    basicstyle=\ttfamily\small, 
    keywordstyle=\color{blue}, 
    commentstyle=\color{gray}\textit, 
    stringstyle=\color{orange}, 
    showstringspaces=false, 
    breaklines=true, 
    frame=tb, 
    framerule=0.5pt, 
    rulecolor=\color{black}, 
    moredelim=[is][\color{mygreen}]{^}{^},
}

\begin{document}

\copyrightyear{2025}
\copyrightclause{Copyright for this paper by its authors.
  Use permitted under Creative Commons License Attribution 4.0
  International (CC BY 4.0).}

\title{\centering{Formal Methods Meets Readability: Auto-Documenting JML Java Code}}


\author{Juan Carlos Recio Abad}[
orcid=0009-0001-0831-2805,
email=jcrecio@uma.es]
\author{Rub\'en Saborido}[
orcid=0000-0002-0944-5941,
email=rsain@uma.es]
\author{Francisco Chicano}[
orcid=0000-0003-1259-2990,
email=chicano@uma.es]
\address{ITIS Software, University of M\'alaga}



\cortext[1]{Corresponding author.}
\fntext[1]{These authors contributed equally.}

\begin{abstract}
This paper investigates whether formal specifications using Java Modeling Language (JML) can enhance the quality of Large Language Model (LLM)-generated Javadocs. While LLMs excel at producing documentation from code alone, we hypothesize that incorporating formally verified invariants yields more complete and accurate results. We present a systematic comparison of documentation generated from JML-annotated and non-annotated Java classes, evaluating quality through both automated metrics and expert analysis.
Our findings demonstrate that JML significantly improves class-level documentation completeness, with more moderate gains at the method level. Formal specifications prove particularly effective in capturing complex class invariants and design contracts that are frequently overlooked in code-only documentation. A threshold effect emerges, where the benefits of JML become more pronounced for classes with richer sets of invariants.
While JML enhances specification coverage, its impact on core descriptive quality is limited, suggesting that formal specifications primarily ensure comprehensive coverage rather than fundamentally altering implementation descriptions. These results offer actionable insights for software teams adopting formal methods in documentation workflows, highlighting scenarios where JML provides clear advantages. The study contributes to AI-assisted software documentation research by demonstrating how formal methods and LLMs can synergistically improve documentation quality.
\end{abstract}


\conference{}
\maketitle

\section{Introduction}
The challenge of maintaining accurate documentation that remains updated with evolving codebases has led researchers to explore automation solutions through Large Language Models (LLMs)~\cite{allamanis2018survey}\cite{chen2021evaluating}. While these models demonstrate remarkable capabilities in generating human-readable text, code and documentation, their effectiveness depends critically on the contextual information available during documentation generation~\cite{liu-etal-2024-lost}. 
Current approaches using LLMs for documentation generation often struggle with inferring implicit program invariants and behavioral contracts directly from implementation code~\cite{zhang2023large}. We hypothesize that Java Modelling Language (JML)~\cite{leavens2006jml} annotations provide structured semantic constraints that enable LLMs to produce more precise, contract-aware documentation that better reflects actual program behavior. Our work examines this hypothesis through a systematic comparison of documentation generated under two distinct conditions: first, from JML-annotated Java classes containing formally verified invariants, and second, from identical Java classes without any specification annotations.

\section{Methodology}
\begin{figure}[ht]
\centering
\includegraphics[width=315pt]{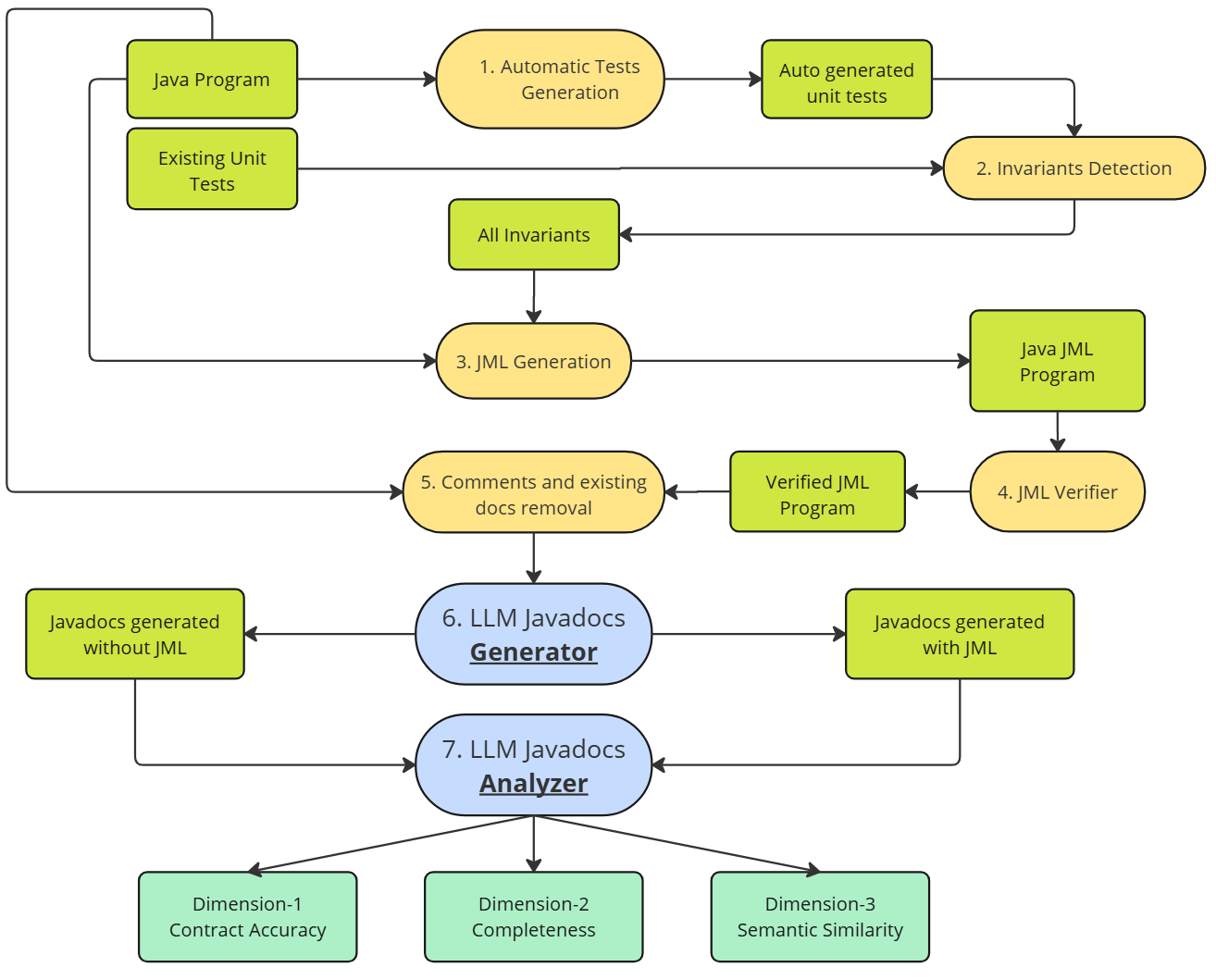}
\caption{Experimental workflow comparing JML-enhanced vs baseline documentation generation}
\label{fig:methodology}
\end{figure}

We establish a controlled experiment using GitHub Java repositories which are highly-starred, with a good test coverage, large number of commits, dozens or hundreds of contributors and regular releases. For each repository, we select a subset of subject classes. We use these classes to generate formal specifications through a multi-stage process, beginning with test generation using EvoSuite~\cite{evosuite} to augment test suite coverage. Subsequent invariant detection employs the Daikon Tool~\cite{Ernst2007Daikon} to identify probable invariants through systematic test execution. This steps benefits from the previous test coverage augmentation as it can help to discover more invariants due to a massive input data instrumentation. The process then annotates detected invariants as JML specifications verified through the KeY theorem prover~\cite{Beckert2007VerificationOO}, followed by removal of unverifiable invariants to retain only formally validated JML annotations.
This rigorous procedure produces two distinct versions of each subject class: a JML-enhanced version containing the original code augmented with verified JML annotations, and a baseline version consisting solely of the original code without any annotations. Documentation generation for both variants utilizes identical LLM configurations designated as the \emph{generator}, maintaining consistent temperature~\cite{sutskever2014sequencesequencelearningneural} settings across all trials. We employ matching prompt templates requesting standard Javadoc elements, while ensuring the context window contains complete class implementations. We remove all existing comments before the generation to prevent any bias.
For objective comparison, we utilize a separate LLM designated as the \emph{analyzer} to evaluate documentation quality across three critical dimensions. The first dimension assesses contract accuracy through quantitative measurement of JML-specified invariants correctly reflected in documentation. Implementation alignment forms the second dimension, calculated via semantic similarity metrics between generated documentation and actual code functionality. Completeness constitutes the third dimension, evaluating coverage of method behaviors, parameter constraints, and return value specifications against ground truth verification results. The diagram ~\ref{fig:methodology} shows the workflow defined by the methodology.

\section{Experiment}
Our evaluation framework analyzes seven medium-complexity Java classes  selected from open-source projects~\cite{google_guava} \cite{google_copybara_2024} \cite{google_gson_2.10.1} 
\cite{facebook_soloader} \cite{oracle_weblogic_monitoring_exporter} meeting our methodology criteria. For each subject class, we generate paired documentation sets consisting of one version derived from JML-annotated code and another from raw code. The \emph{analyzer} LLM evaluates both documentation variants against ground truth specifications independently verified through the KeY Project. For the \emph{generator} we chose Grok-3~\cite{grok2023} as LLM. Regarding preliminary experiments, this model excels at generating Javadocs from Java code, particularly when the code includes JML annotations. Alternatively, for the \emph{analyzer} we opted for Claude Sonnet 3.7~\cite{anthropic2024claude}. It stands out in analyzing and evaluating Javadocs. It possesses strong analytical capabilities that allow it to assess documentation quality through multiple perspectives. Its natural language understanding helps evaluate how effectively technical concepts are communicated in plain language, while its analytical framework enables systematic comparison of documentation against established standards or requirements. 
Once we receive the responses from the \emph{generator}, we send the Javadocs to the \emph{analyzer} for analysis comparison several times, in order to get the average results and avoid misleading results due to the stochastic nature of the model (even with the temperature set to zero).
Our experimental evaluation reveals nuanced insights into the impact of JML annotations on documentation quality across different granularity levels. At the class documentation level, JML-enhanced generation achieves a 92\% completeness score, compared to 85\% for non-JML documentation. Method-level documentation shows more modest gains, with JML-derived documentation reaching 92\% completeness versus 87\% for baseline generation.
The semantic similarity analysis presents an interesting counterpoint to these completeness metrics. Both JML-enhanced and baseline documentation achieve comparable 85\% average similarity when describing identical functionality, suggesting that while JML annotations improve coverage of formal specifications, they may not substantially alter the core descriptive quality of the generated text. This finding implies that the primary value of JML annotations lies in their ability to ensure comprehensive inclusion of formal constraints, rather than fundamentally changing how implementations are described.
We observe that JML annotations contribute to documentation completeness without substantially increasing verbosity, as the model synthesizes formal constraints into concise yet comprehensive descriptions. Listing \ref{lst:listing1} shows a method extracted from the Guava Google project that hugely benefits from the JML annotations. What the method does is trivial, however the class context does not capture at all the nuances that JML annotations do. For this reason, in this case the contract accuracy for No-JML generated documentation is 54\% versus a 94\% from the JML version, with a 39\% semantic similarity to describe the functionality.

\begin{lstlisting}[caption={Java method with huge benefits on generated documentation due to the annotated invariants. Invariants are simplified for space optimization.},captionpos=b,style=tfmstyle,basicstyle=\scriptsize,label={lst:listing1}]
  /*@ requires ipString != null;
  @ ensures IPV4_DELIMITER_MATCHER == \old(IPV4_DELIMITER_MATCHER);
  @ ensures IPV4_DELIMITER_MATCHER.getClass().getName() == \old(IPV4_DELIMITER_MATCHER.getClass().getName());
  @ ensures IPV4_DELIMITER_MATCHER.getClass().getName() == \old(IPV6_DELIMITER_MATCHER.getClass().getName());
  @ ensures IPV6_DELIMITER_MATCHER == \old(IPV6_DELIMITER_MATCHER);
  @ ensures LOOPBACK4 == \old(LOOPBACK4);
  @ ensures ANY4 == \old(ANY4);
  @ ensures ipString.toString().equals(\old(ipString.toString()));*/
  public static boolean isInetAddress(String ipString) {
    return ipStringToBytes(ipString, null) != null;
  }
\end{lstlisting}

\begin{table}[ht]
\footnotesize
\centering
\caption{Lowest score (Row 1) and highest score (Row 3) for Javadocs obtained in the entire experiment for the class \emph{InetAddresses}. Row 2 shows an average analysis between two generated Javadocs for No-JML and JML. Contract Accuracy is represented by D1, Completness is represented by D2 and Similarity between Javadocs is represented by D3.}
\label{tab:method_comparison}
\begin{tabular}{lccccccc}
\toprule
 & \multicolumn{2}{c}{\textbf{NO-JML}} & \multicolumn{2}{c}{\textbf{JML}} & & & \\
\cmidrule(lr){2-3} \cmidrule(lr){4-5}
\textbf{Method} & \textbf{D1 (\%)} & \textbf{D2 (\%)} & \textbf{D1 (\%)} & \textbf{D2 (\%)} & \textbf{D3 (\%)} & \textbf{\# Invariants} \\ 
\midrule
\texttt{isInetAddress}         & 54 & 62 & 82 & 94 & 39 & 9 \\
\texttt{textToNumericFormatV4} & 81 & 79 & 84 & 90  & 85 & 15  \\
\texttt{textToNumericFormatV6} & 82 & 85 & 87 & 92 & 87 & 32 \\
\bottomrule
\end{tabular}
\end{table}

\section{Conclusion}
This study demonstrates that JML annotations enhance the quality of automatically generated Javadocs, particularly for class-level documentation and methods with complex invariants. Our experiments reveal that formal specifications help bridge the gap between implementation and documentation by providing structural guidance to LLMs, ensuring comprehensive coverage of contracts and constraints. While method-level documentation shows more modest improvements, the largest gains occur in classes with rich invariants, where JML-derived documentation achieves near-complete specification coverage.
The consistent semantic similarity between JML and non-JML documentation suggests that LLMs generate similarly fluent descriptions regardless of annotation status. However, JML acts as a safeguard against omissions, particularly for subtle preconditions, postconditions, and class invariants. This trade-off between completeness and conciseness warrants further investigation, especially in real-world development scenarios where documentation usability is critical. Future work should explore hybrid approaches that combine JML specifications with natural language augmentation to capture both formal and informal aspects of software behavior. Additionally, more studies with developers could validate if the improved metrics translate to tangible productivity gains in maintenance tasks.
\bibliography{sample-ceur}

\end{document}